\documentclass[showpacs,aps,prl,reprint,superscriptaddress,preprintnumbers]{revtex4-1}
\usepackage{ulem}
\usepackage{amsmath}
\usepackage{amssymb}
\usepackage{graphicx}
\usepackage{dcolumn}
\usepackage{color}
\usepackage{xcolor}
\usepackage{endnotes}
\usepackage{bm}
\usepackage{epsfig}
\usepackage{array}

\newcommand{\SrIr}{Sr$_{3}$Ir$_4$Sn$_{13}$}
\newcommand{\SrRh}{Sr$_{3}$Rh$_4$Sn$_{13}$}

\newcommand{\CaSrRhx}{(Ca$_{x}$Sr$_{1-x}$)$_3$Rh$_4$Sn$_{13}$}

\newcommand{\CaSrIrx}{(Ca$_{x}$Sr$_{1-x}$)$_3$Ir$_4$Sn$_{13}$}

\newcommand{\CaIr}{Ca$_{3}$Ir$_4$Sn$_{13}$}

\begin{document}

\title{Peak in the critical current density in (Ca$_{x}$Sr$_{1-x}$)$_3$Rh$_4$Sn$_{13}$ tuned towards\\ the structural quantum critical point }

\author{Xinyou Liu}
\author{Wei Zhang}
\author{Kwing To Lai}
\affiliation{Department of Physics, The Chinese University of Hong Kong, Shatin, Hong Kong SAR, China}
\author{K.~Moriyama}
\affiliation{Department of Chemistry, Kyoto University, Kyoto 606-8502, Japan}
\author{J.~L.~Tallon}
\affiliation{Robinson Institute, Victoria University of Wellington, P. O. Box 600, Wellington, New Zealand}
\author{K.~Yoshimura}
\affiliation{Department of Chemistry, Kyoto University, Kyoto 606-8502, Japan}
\author{Swee~K.~Goh}
\email[]{skgoh@cuhk.edu.hk}
\affiliation{Department of Physics, The Chinese University of Hong Kong, Shatin, Hong Kong SAR, China}

\date{\today}

\begin{abstract}
(Ca$_{x}$Sr$_{1-x}$)$_3$Rh$_4$Sn$_{13}$ is a rare system that has been shown to display an interesting interplay between structural quantum criticality and superconductivity. 
A putative structural quantum critical point, which is hidden beneath a broad superconducting dome, is believed to give rise to optimized superconducting properties in (Ca$_{x}$Sr$_{1-x}$)$_3$Rh$_4$Sn$_{13}$. However, the presence of the superconducting dome itself hinders the examination of the quantum critical point through electrical transport, as the transport coefficients vanish in the superconducting state. Here, we use critical current density to 
explore within the superconducting dome. Our measurements reveal a large enhancement of the critical current density at the zero-temperature limit when the system is tuned towards the structural quantum critical point. 
\end{abstract}


\maketitle


A quantum critical point (QCP) is a location in parameter space where a second-order quantum phase transition occurs at zero temperature. Its prominence in condensed matter physics is tied to the observation of superconductivity in a large variety of materials in the vicinity of the putative QCP \cite{Paschen2021,Matthias2018,Keimer2017,Paglione2010,Dai2015,Gupta2021,Frachet2020,Gupta2021,Frachet2020,Tallon1999,Cho2018,Hashimoto2012,Analytis2014,Putzke2014,Shibauchi2014,Gegenwart2008,Park2006,Mathur1998,Gruner2017}. Examples where 
the presence of the QCP has been proposed include cuprates  \cite{Tallon1999,Gupta2021,Frachet2020}, iron pnictides \cite{Analytis2014,Hashimoto2012,Shibauchi2014,Putzke2014}, charge density wave (CDW) systems \cite{Gruner2017,Cho2018}, and heavy-fermion systems \cite{Mathur1998,Park2006,Gegenwart2008}, in which a dome-shaped dependence of the superconducting transition temperature $T_c$ is found. 
However, such a superconducting dome naturally masks the QCP. Usually, to reveal the exact pressure or composition where the QCP is located requires the extrapolation of the relevant phase transition temperature to zero, or the application of a strong magnetic field to suppress superconductivity. To probe the QCP directly, a physical parameter which does not vanish in the superconducting state is required. 

Critical current density $J_c$ is just such a parameter that does not vanish inside the superconducting dome. $J_c$ in the zero-temperature limit, $J_c(0)$, is particularly suitable for detecting the presence of a QCP, as it is a singular point which exists precisely at the absolute zero temperature. In the heavy-fermion system Sn-doped CeRhIn$_5$, an antiferromagnetic QCP can be reached by applying a hydrostatic pressure of 1.35~GPa \cite{Seo2015}. Indeed, at the antiferromagnetic QCP, an enhancement of $J_c$ was reported \cite{Jung2018}, and the location of the QCP is consistent with that determined by other physical properties of Sn-doped CeRhIn$_5$ through extrapolations of the antiferromagnetic phase transition from finite temperatures \cite{Seo2015,Ferreira2008}. In high-$T_c$ cuprates, $J_c(0)$ has also been shown to peak at the critical doping \cite{Naamneh2014,Talantsev2014}. 
 Therefore, $J_c(0)$ offers a powerful means to probe the true ground-state property of the system along the $T=0$ axis inside a superconducting dome.

To establish the extent to which $J_c(0)$ can detect a QCP, it is important to look beyond a magnetic QCP, because a nonmagnetic QCP can also support the optimization of superconductivity in an analogous manner. Quasiskutterudite  ``Remeika" compounds M$_{3}$T$_{4}$Sn$_{13}$, where M is an  alkaline-earth metal and T is a transition metal, are promising systems to explore this relationship~\cite{Remeika1980,Kase2011,Goh2015,Yu2015,Cheung2018,Lue2016,Klintberg2012,Chen2016,Veiga2020,Terasaki2021,Biswas2014,Biswas2015,Krenkel2021,Zhou2012,Sarkar2015,Yang2010,Luo2018,Hu2017,Kaneko2019}. \SrRh\ is a cubic system which shows a second-order structural transition at $T^*=138$~K and a superconducting transition at $T_c=4.2$~K \cite{Kase2011,Goh2015,Yu2015,Cheung2018,Lue2016}. Upon the substitution of Ca to form the series \CaSrRhx, $T^*$ can be suppressed rapidly, and it extrapolates to 0~K at $x_c\sim0.9$, giving rise to the proposal of a structural QCP at $x_c$ \cite{Goh2015,Yu2015,Cheung2018,Terasaki2021,Krenkel2021}. Concomitant with the suppression of $T^*$, $T_c$ follows a dome-shaped dependence on $x$ with a maximum $T_c$ of $\sim$7.9~K at the calcium-rich end of the series. Further experimental support for the structural QCP at $x_c$ includes a linear-in-$T$ resistivity from $T_c$ to $\sim$40~K, a minimum Debye temperature, and a complete softening of the relevant phonon mode at the {\bf M} point of the Brillouin zone in the zero-temperature limit \cite{Goh2015,Yu2015,Cheung2018}. Moreover, strong-coupling superconductivity is stabilized near $x_c$, as benchmarked by large values of  $2\Delta_{sc}/k_BT_c$ and $\Delta C/\gamma T_c$ that are both beyond the BCS weak-coupling limit \cite{Goh2015,Yu2015,Cheung2018,Terasaki2021}. The related series \CaSrIrx\ behaves similarly \cite{Klintberg2012}, but $T^*$ only extrapolates to 0~K when \CaIr\ is subjected to a hydrostatic pressure of 1.8~GPa \cite{Klintberg2012,Biswas2015}. Thus \CaSrRhx\ offers a platform to study a structural QCP without the need of applying pressure.

In this manuscript, we report our $J_c$ data for the entire substitution series \CaSrRhx, straddling the putative structural QCP. With the availability of $J_c$ data down to 50~mK a reliable extraction of $J_c(0)$ for all samples can be achieved. Our $J_c(0)$ reveals a significant enhancement when the system is tuned to $x_c$, offering compelling evidence for the existence of a structural QCP in \CaSrRhx.

\begin{figure}[!t]\centering
      \resizebox{9cm}{!}{
              \includegraphics{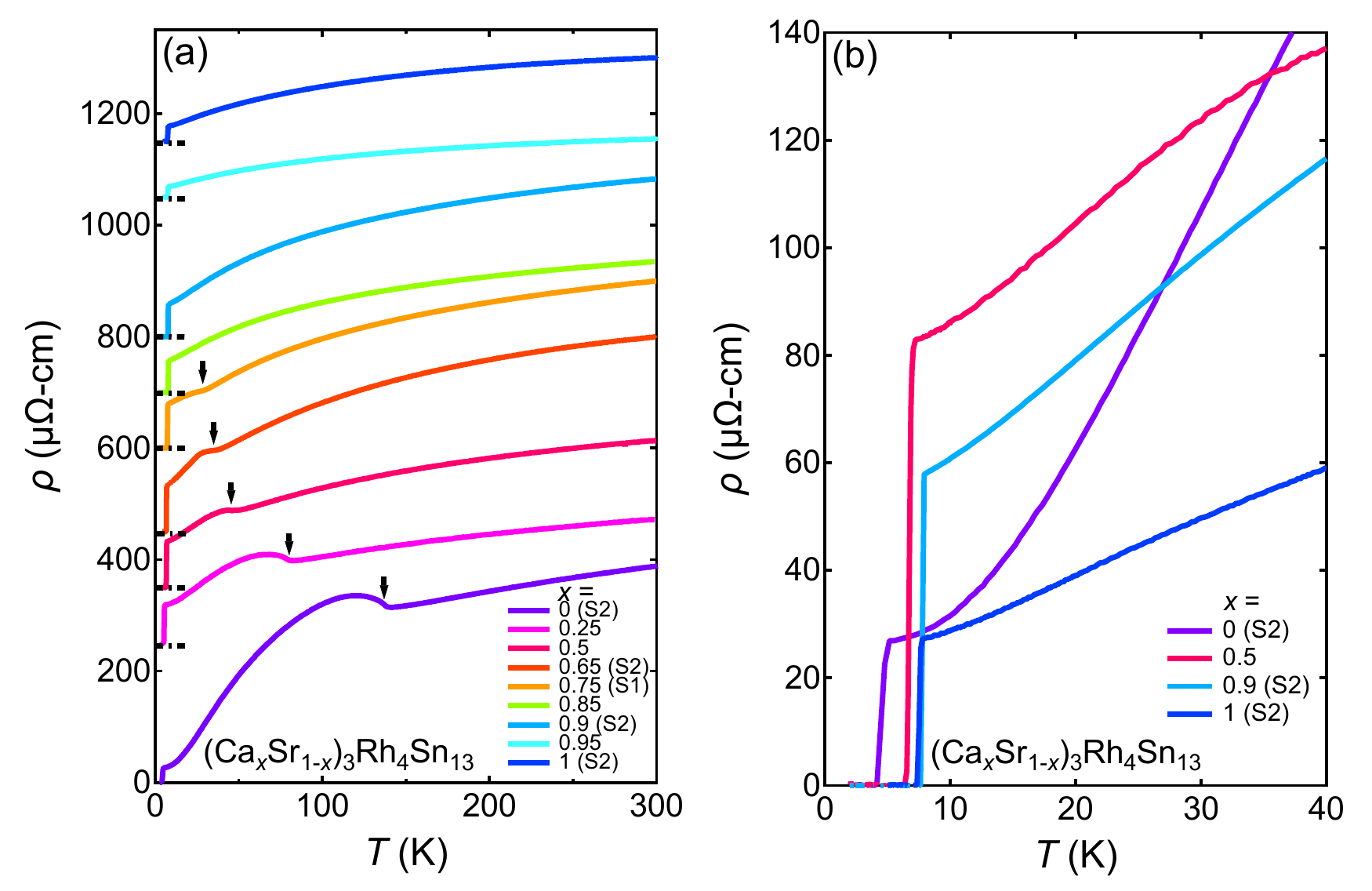}}                				
              \caption{\label{fig1} (a) Temperature dependence of the electrical resistivity for \CaSrRhx\ across the entire substitution series. The traces are offset vertically for clarity. The arrows indicate $T^*$, the temperature at which the second-order structural transition occurs. Zero resistivity for each curve is marked by the horizontal dashed line. (b) The low-temperature zoom-in of $\rho(T)$ for $x = 0$, $x = 0.5$, $x=x_c = 0.9$, and $x$ = 1 samples. }
\end{figure}

Single crystals of \CaSrRhx\ were synthesized by the Sn-flux method as described elsewhere \cite{Yang2010}. Electrical transport measurements were carried out in a Physical Property Measurement System (PPMS) from Quantum Design and in a dilution fridge manufactured by Bluefors. Temperature-dependent electrical resistivity was measured with a conventional four-probe configuration. Dupont 6838 silver epoxy and gold wires with diameters of 10~$\mu$m or 25~$\mu$m were used for making the electrical contacts. 
$I$-$V$ curves were measured using the same contacts by a Keithley 2182A nanovoltmeter combined with a Keithley 6221 current source in the pulsed delta mode. The duration of the pulsed current was 11~ms, and the pulse repetition time was 0.1~s. 
These pulse sequence parameters not only minimize Joule heating, but also ensure a practical measurement time \cite{SUPP}.


Figure~\ref{fig1}(a) shows the temperature dependence of the electrical resistivity $\rho(T)$ of {(Ca$_x$Sr$_{1-x}$)$_3$Rh$_4$Sn$_{13}$} from $x$ = 0 to $x$ = 1 upon cooling from 300~K. All $\rho(T)$ traces are progressively offset vertically for clarity, with the zero resistivity for each trace marked by the horizontal dashed line. A prominent anomaly in $\rho(T)$ is recorded in the $x$ = 0 sample at $T^*$ = 137~K, which corresponds to a second-order structural phase transition from a high-temperature crystal structure with a space group $Pm\bar{3}n$ to a low-temperature structure $I4\bar{3}d$ below $T^*$. To determine $T^*$, we use the minimum in the first derivative of resistivity, and $T^*$ decreases rapidly with an increasing Ca fraction $x$. In the $x$ = 0.75 sample, $T^*$ is 26.6~K, but it becomes difficult to trace $T^*$ further with higher $x$ values. Figure~\ref{fig1}(b) shows the low-temperature region of $\rho(T)$. The superconducting transition temperature $T_c$, defined as the midpoint of the transition, experiences a significant enhancement from 4.5~K in the $x = 0$ sample to 7.9~K in the $x = 1$ sample. Furthermore, a quasilinear $\rho(T)$ is observed from $T_c$ to 40~K near $x=x_c = 0.9$, which has been attributed to the softening of phonon modes at the structural QCP. The curvature of $\rho(T)$ is gradually recovered when $x$ moves away from $x_c$. These observations are consistent with previous reports \cite{Goh2015,Yu2015,Cheung2018}.  

\begin{figure}[!t]\centering
       \resizebox{8.5cm}{!}{
              \includegraphics{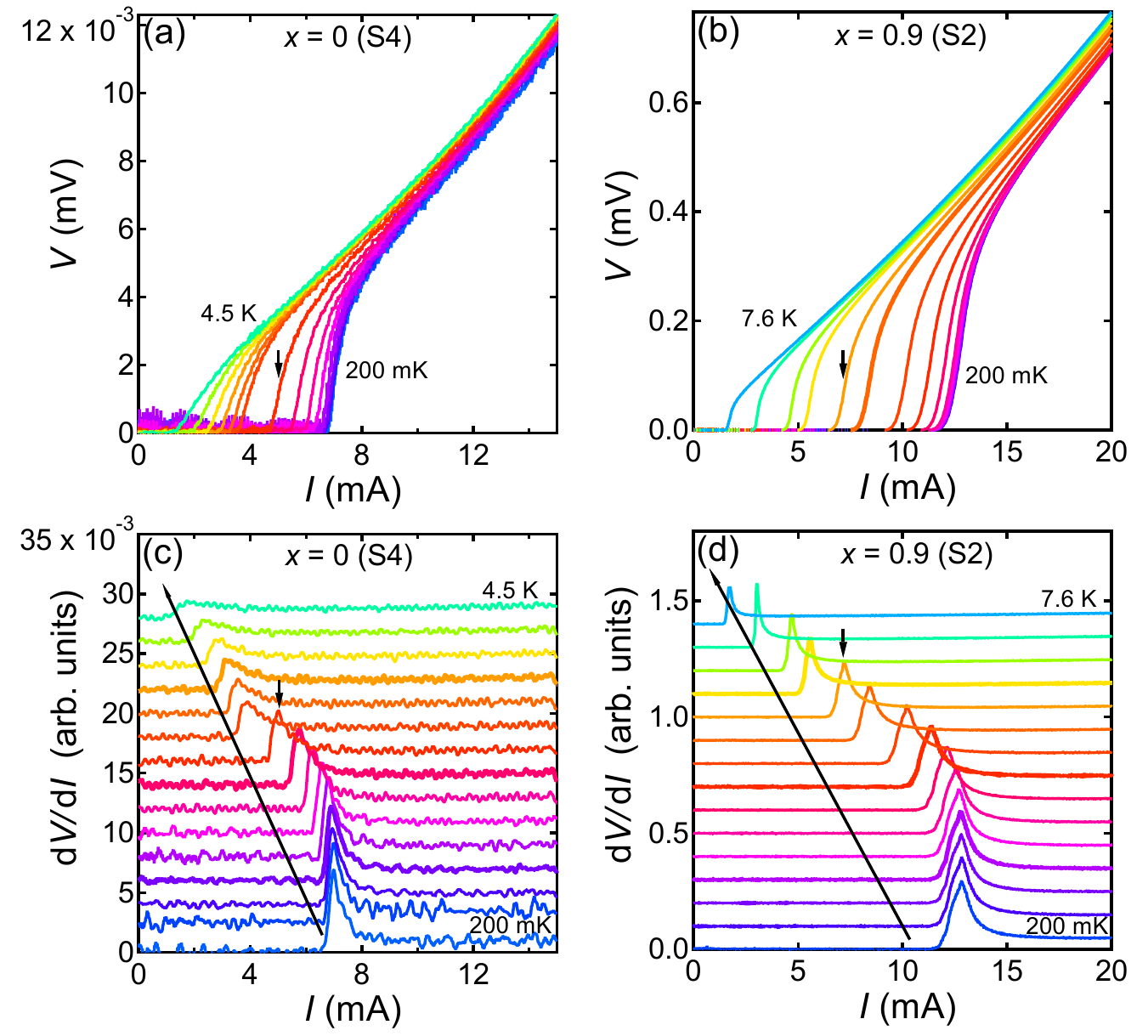}}                				
              \caption{\label{Fig2} $V$-$I$ characteristics at various temperatures for (a) $x = 0$ (sample~S4) and (b) $x_c = 0.9$ (sample~S2). The calculated first derivative of $V(I)$ (d$V$/d$I$) for (c) $x = 0$ and (d) $x_c = 0.9$. The long arrows show the direction of an increasing temperature. The peak positions in d$V$/d$I$ define the corresponding $I_c$. The vertical arrows illustrate the $I_c$ at 3.5~K for $x=0$, and at 6.5~K for $x_c$.}
\end{figure}

We proceed to measure the critical current $I_c$ for 13 samples spanning the entire substitution series. For $x$ = 0, 0.65, 0.9, and 1, two samples from the same batch have been measured for each composition (see Supplemental Material \cite{SUPP}). For each sample, we measure the voltage at a fixed temperature when a pulsed current is applied perpendicular to the cross section of the specimen with current leads covering the whole width.
$V$-$I$ curves at various temperatures for $x = 0$ (sample~S4) are shown in Fig.~\ref{Fig2}(a). For a typical trace, a drastic increase of the voltage from zero is recorded when the current exceeds a threshold, indicating a recovery from the superconducting state to the normal state. 
To quantify the value of $I_c$, we calculate the first derivative of $V(I)$. Figure~\ref{Fig2}(c) displays d$V$/d$I$ versus $I$ for the same $x=0$ sample, and the peak positions are defined as the corresponding $I_c$. When the temperature is reduced, $I_c$ increases significantly and saturates at the zero-temperature limit. As we will argue later, this saturation is consistent with a dominant nodeless order parameter. Figures~\ref{Fig2}(b) and (d) show the $V$-$I$ curves and d$V$/d$I$ versus $I$ for $x=x_c=0.9$  (sample~S2), respectively. However, we point out that $I_c$ at the zero-temperature limit, $I_c(0)$, is noticeably higher in the $x=0.9$ sample, despite the fact that the $x=0.9$ sample has a smaller cross-sectional area than that of the $x=0$ sample.
\begin{figure}[!t]\centering
       \resizebox{8.5cm}{!}{
              \includegraphics{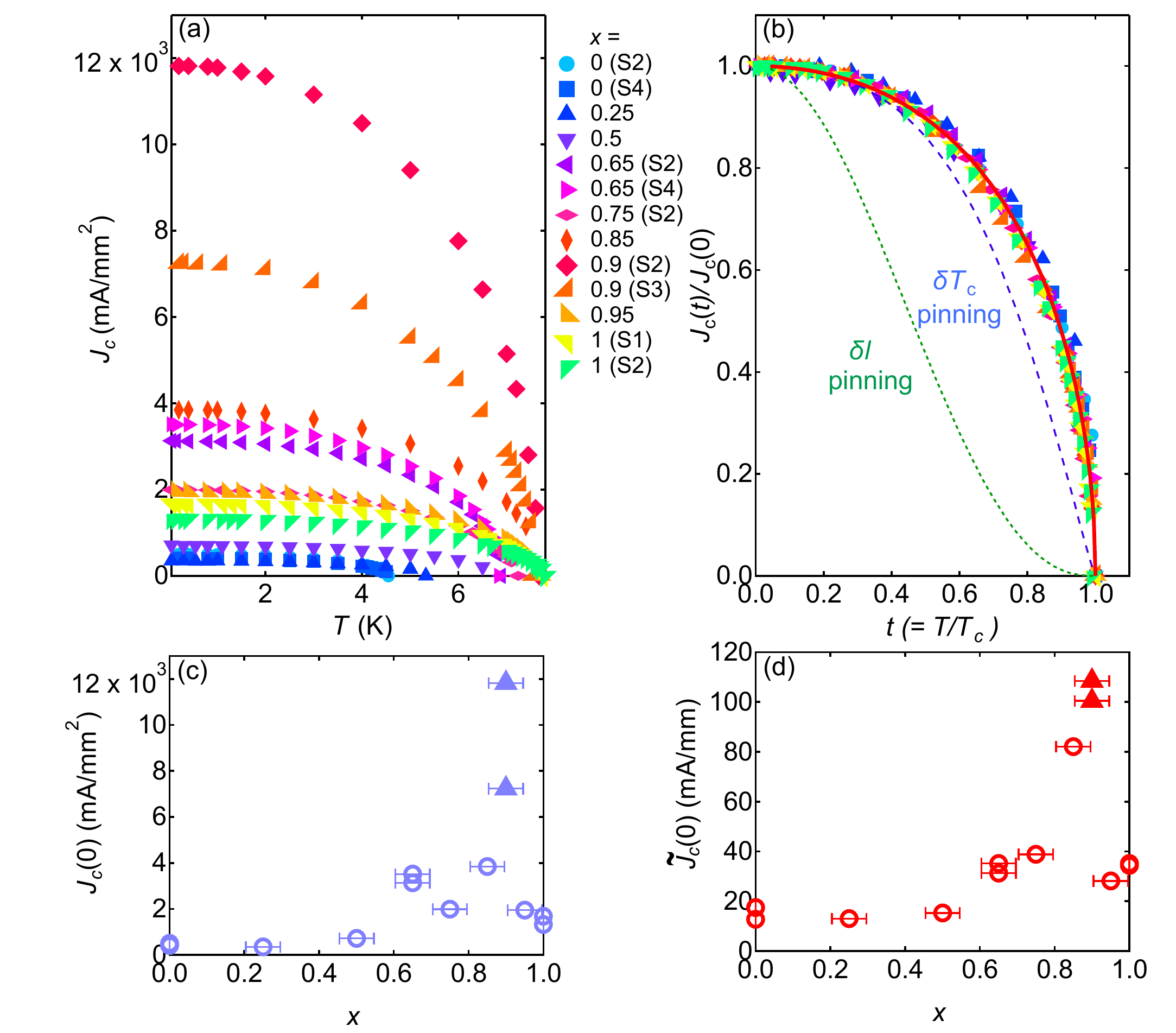}}                				
              \caption{\label{fig3} (a) Temperature dependence of $J_c$ and (b) $J_c$ normalized by $J_c(0)$ as a function of the reduced temperature $t=T/T_c$ for all samples.  The normalized $J_c(t)$ collapse onto a universal curve, which can be described by the red solid curve governed by Eq.(\ref{eqn_pinning}) with $\alpha=0.5$ and $\beta=0.2$. Blue dashed and green dotted lines represent the curve for $\delta T_c$-pinning and $\delta l$-pinning mechanisms, respectively \cite{Griessen1994}. (c) Ca content $x$ dependence of $J_c(0)$ and (d) $\widetilde{J_c}(0)$ for all the samples. The solid triangles in (c) and (d) denote the $J_c(0)$ and $\widetilde{J_c}(0)$ = $J_c(0)\times b$ of the two $x$=0.9 samples. Note that $x$ = 0.75 (sample S2)  is from the same batch as $x$ = 0.75 (sample S1), whose $\rho(T)$ is shown in Fig. 1. The error bars in $x$ are determined following a scheme described in Ref.~\cite{SUPP}.}
\end{figure}

For a sample with a width of $2a$ and a thickness of $2b$, $J_c=I_c/(2a\times 2b)$, as conventionally defined. Figure~\ref{fig3}(a) shows the temperature dependence of $J_c$ for all 13 samples. We note the discrepancy in the $J_c$ values for the two $x=0.9$ samples, which we are able to resolve below. Nevertheless, the overall temperature dependencies for all samples are similar, despite their different $T_c$. To visualize the similarity, we plot in Fig.~\ref{fig3}(b) $J_c$ normalized by $J_c(0)$ as a function of the reduced temperature $t=T/T_c$. Indeed, all $J_c$ data collapse onto the same curve when plotted in this manner, hinting at a universal mechanism governing the temperature dependence of $J_c$. We find empirically that the normalized $J_c$ can be accurately described by 
\begin{equation}
\frac{J_c(t)}{J_c(0)}=(1-t^2)^\alpha(1+t^2)^\beta,
\label{eqn_pinning}
\end{equation}
with $\alpha=0.5$ and $\beta=0.2$ (red solid curve in Fig.~\ref{fig3}(b)). In fact, Eq.~(\ref{eqn_pinning}) with $(\alpha,\beta)=(7/6,5/6)$ is frequently employed to described $J_c(t)$ when the so-called $\delta T_c$-pinning is effective \cite{Griessen1994}. For $\delta l$-pinning (where $l$ is the mean free path), $(\alpha,\beta)=(5/2,-1/2)$ \cite{Xiang2013}. The curves due to $\delta l$- and $\delta T_c$-pinnings are added to Fig.~\ref{fig3}(b), and both mechanisms are incompatible with our data. This is not surprising because, in the absence of the applied magnetic field, the vortex physics does not play a role. Hence, the pinning mechanisms are irrelevant to our discussion.

We adopt the viewpoint that our $J_c(T)$ reflects, to a certain extent, the temperature dependence of $\lambda^{-2}$, where $\lambda$ is the London penetration depth. This viewpoint emerges from a recent discovery that the self-field critical current, {\it i.e.} a transport $J_c$ without an applied magnetic field, is fundamentally determined by $\lambda$. As demonstrated in Refs.~\cite{Talantsev2015,srTalantsev2017}, the self-field $J_c$ is given by 
\begin{equation}
J_c=\frac{\phi_0}{4\pi\mu_0\lambda^3}\left(\ln\left(\frac{\lambda}{\xi}\right)+0.5\right)\times\frac{\lambda}{b}\tanh\left(\frac{b}{\lambda}\right),
\label{eqn_full}
\end{equation}
where $\phi_0$ is the flux quantum, $\mu_0$ is the vacuum permeability, and $\xi$ is the coherence length. 

For our samples, the thickness $2b$ is of the order of several tens of $\mu {\rm m}$ -- the thinnest sample has $2b=17~\mu{\rm m}$ \cite{SUPP}. On the other hand, $\lambda$ in the zero-temperature limit, $\lambda(0)$, has been reported to be $\sim$290~nm in the related compound \SrIr~ \cite{Biswas2014}. Assuming that $\lambda(0)$ in \CaSrRhx\ is of the same order of magnitude, then $b/\lambda(0)\gg1$. Equation~(\ref{eqn_full}) in the `thick limit' becomes
\begin{equation}
J_c=\frac{\phi_0}{4\pi\mu_0\lambda^2}\left(\ln\left(\frac{\lambda}{\xi}\right)+0.5\right)\times\frac{1}{b}.
\label{eqn_thick}
\end{equation}

Near $T_c$, $\lambda$ diverges and can potentially complicate the analysis, because $\lambda$ will eventually exceed $b$. However, given the thickness of our samples \cite{SUPP}, this would occur very near $T_c$. In the Bardeen-Schrieffer-Cooper weak-coupling limit, $\lambda$ will only reach $10\lambda(0)$ at $T/T_c=0.995$.  Here, our sample thickness is roughly two orders of magnitude larger than $\lambda(0)$. Therefore, the condition $b/\lambda(0)\gg1$ is essentially valid for all samples at all practical $T<T_c$.

From Eq.~(\ref{eqn_thick}), one can immediately see that $J_c$ depends on a sample-specific extrinsic factor (the half sample thickness $b$), while all other factors are either constants or material parameters. This provides a hint to reconcile the drastic difference between the $J_c$ values of the two $x=0.9$ samples mentioned above: The difference is simply due to the different thicknesses. To illustrate the validity of this hypothesis, the critical current density at the zero-temperature limit of all samples before and after taking this factor into account is shown in Figs.~\ref{fig3}(c) and (d). Indeed, when the thickness is taken into account, $\widetilde{J_c}(0)=J_c(0)\times b$ of the two $x=0.9$ samples become much closer: The $\widetilde{J_c}(0)$ values are $\pm3.74\%$ from the mean value, while before correcting for the thickness, the $J_c(0)$ values are $\pm24.16\%$ from the mean value (see the solid triangles in Figs.~\ref{fig3}(c) and (d)). In fact, $\widetilde{J_c}(0)$ over the entire range of the Ca content $x$ becomes less scattered than that of $J_c(0)$, as can be seen by comparing Fig.~\ref{fig3}(d) with Fig.~\ref{fig3}(c). To further verify the $1/b$ dependence of $J_c$, we measure another piece of $x=0.5$ crystal, which we polish to three different thicknesses, all in the thick limit. As presented in the Supplemental Material~\cite{SUPP}, this measurement with the same starting crystal confirms the applicability of $1/b$ scaling of $J_c$.

Despite the confirmation of the $1/b$ scaling, the measured $J_c$ values are lower than the $J_c$ values calculated from Eq.~(\ref{eqn_thick}). Take, for instance, $x=0.9$~(sample S2) with $2b=17~\mu$m. Assuming $\lambda(0)\sim290$~nm and $\lambda/\xi\sim30$~\cite{Biswas2014}, the calculated $J_c$ is $715 \times 10^3$~mA/{mm}$^{2}$, about 60 times larger than the measured values ($12 \times 10^3$~mA/{mm}$^{2}$). One possibility is the existence of weakly linked superconducting domains in our samples. Thus, the cross-sectional area through which the transport current flows used for converting from $I_c$ to $J_c$ is too large. In the presence of weak links, $J_c^{\rm WL}=\hbar/(2\mu_0de\lambda^2$), where $d$ is some characteristic length scale associated with the weak-link morphology~\cite{Tallon2006,Ambegaokar1963}. Therefore, $J_c^{\rm WL}$ is also proportional to $\lambda^{-2}$.
If the domains are sufficiently large (such that $b\gg\lambda$) and close to each other (so that $d$ is exceedingly small), $J_c^{\rm WL}$ could be too large to be the current limiting mechanism, and the mechanism giving rise to Eq.~(\ref{eqn_thick}) prevails.
This probably explains the observed $1/b$ dependence, and without the precise knowledge of the domain size, the change in $J_c$ can only be associated with the change in $\lambda^{-2}$ with an indeterminate proportionality factor.

The realization that the self-field $J_c$ is proportional to $\lambda^{-2}$ offers a means to evaluate the superconducting gap. Since the ratio $\lambda/\xi$ enters the logarithm, whose temperature dependence is weak, $J_c(t)/J_c(0)$ is dominated by $\lambda^{-2}(t)/\lambda^{-2}(0)$, which reflects the temperature dependence of the superfluid density. As Fig.~\ref{fig3}(b) depicts, $J_c(t)/J_c(0)$ has a weak temperature dependence at low temperatures. Below $t=0.2$, $J_c(t)/J_c(0)$ only deviates from the zero-temperature value by at most $1.5\%$. Therefore, this rapid saturation of $J_c$ as temperature decreases suggests a nodeless order parameter, consistent with the consensus on this material system \cite{Luo2018,Krenkel2021,Yu2015,Kase2011,Zhou2012,Sarkar2015}. 

\begin{figure}[!t]\centering
   \resizebox{8.5cm}{!}{
           \includegraphics{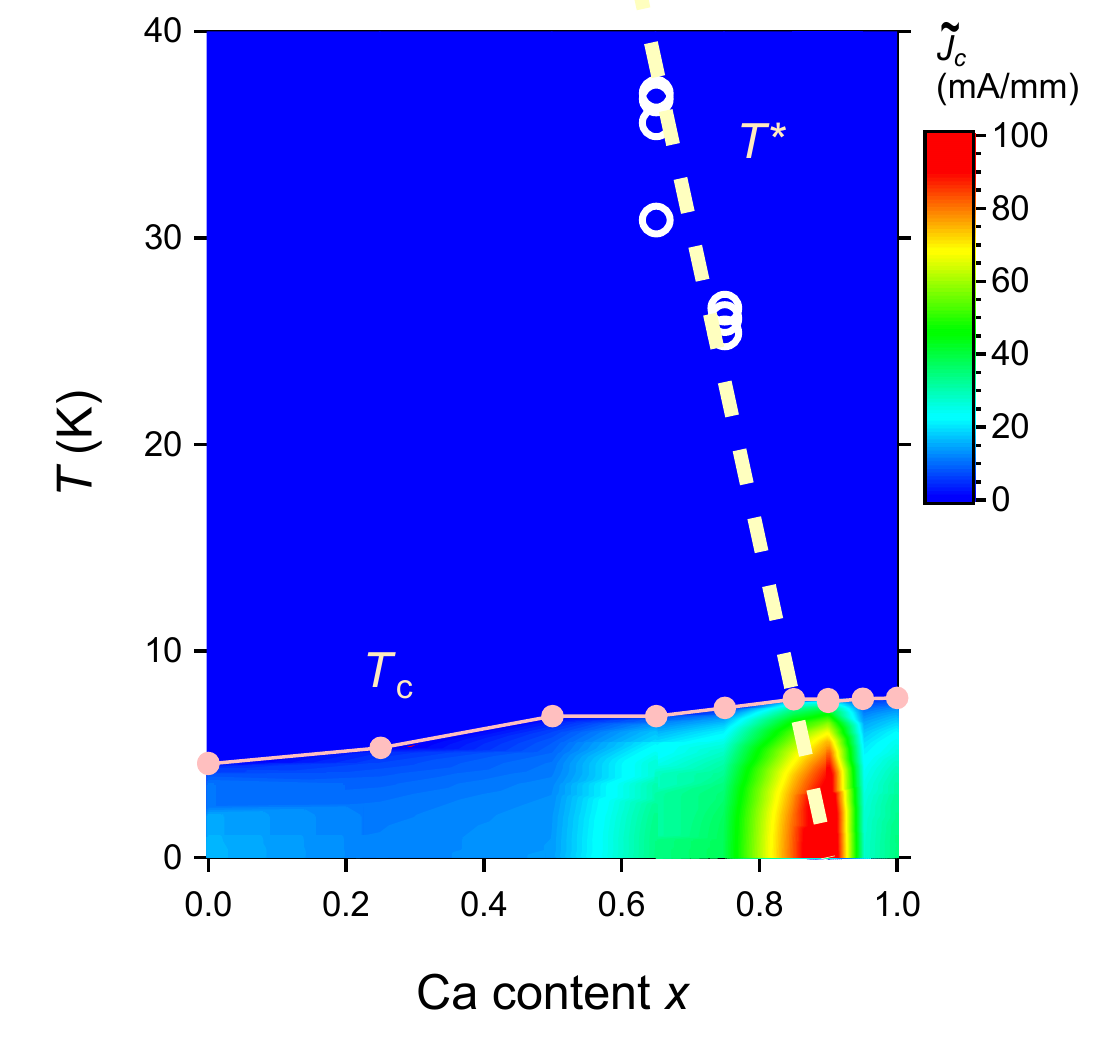}}                				
          \caption{\label{fig4} A contour map of $\widetilde{J_c}$ constructed using our experimental data superimposed on the $T$-$x$ phase diagram. Open circles denote $T^*$, while solid circles denote $T_c$. The dashed line is the $T^*(x)$ from Ref.~\cite{Goh2015}. }
\end{figure} 

We now present our $J_c$ data in relation to the $T$-$x$ phase diagram, compactly summarized in Fig.~\ref{fig4}. The colour map indicates the magnitude of $\widetilde{J_c}$. Again, we use $\widetilde{J_c}$ to eliminate the effect of the sample thickness. In the Ca-rich end, $\widetilde{J_c}$ is generally larger. At $T=0$, $\widetilde{J_c}(0)$ is gradually enhanced with an increasing $x$, and it peaks at $x=0.9$, before decreasing again. Note that the enhancement in $\widetilde{J_c}(0)$ cannot be attributed simply to the enhancement in $T_c$ -- moving from $x=0.5$ to $x=0.9$, $T_c$ only increases by 10\% while $\widetilde{J_c}(0)$ rises by 400$-$500\%.
Incidentally, the maximum $\widetilde{J_c}(0)$ occurs close to the composition where $T^*(x)$ extrapolates to 0~K. Therefore our data provide compelling evidence for the existence of a structural quantum critical point hidden beneath the superconducting dome. Taking together the transport, thermodynamic, and inelastic scattering results \cite{Goh2015,Yu2015,Cheung2018}, the existence of a structural quantum critical point in \CaSrRhx\ at around $x=0.9$ is now firmly established.

Our finding of a peak in the critical current is reminiscent of the recent discovery by Jung {\it et al.}~\cite{Jung2018} in Ce-based heavy-fermion compounds, in which a magnetic QCP hidden beneath the superconducting phase was also revealed by an enhancement of the critical current.
In addition, the hole-doped high-$T_c$ cuprate superconductor Y$_{0.8}$Ca$_{0.2}$Ba$_2$Cu$_3$O$_y$ also features a sharp peak in $J_c$ at the critical hole doping $p=0.19$~\cite{Talantsev2014, Talantsev2015}, where a quantum critical point was proposed to be 
located. In \CaSrRhx, the peak appears sharper than in both heavy-fermion compounds and some cuprates~\cite{Talantsev2014,Talantsev2015,Jung2018}. Other examples of a sharp quantum-critical like peak in $J_c$ do in fact exist for cuprates \cite{Tallon1999,Naamneh2014}. Such a strong similarity across three very distinct classes of superconductors deserves further investigation.
Taken together, these works suggest that the critical current is indeed a suitable diagnostic probe to locate the quantum critical point masked by the superconducting dome. As a future direction, the measurement of $J_c$ in a magnetic field will enable the discussion of pinning mechanisms, potentially revealing new physics associated with the suppression of structural order. For instance, magnetization measurements in $\text{Ba}{({\text{Fe}}_{1\ensuremath{-}x}{\text{Co}}_{x})}_{2}{\text{As}}_{2}$ also show a peak in $J_c$ close to the optimum doping~\cite{Prozorov2009}: The in-field measurements led to the proposal that the $J_c$ peak is related to pinning on structural domains. 

The peak in the critical current density would imply a dip in the London penetration depth (see Eqs.~(\ref{eqn_full}) and (\ref{eqn_thick})). This is counter-intuitive for our system. When the electron-phonon coupling is taken into account, the London penetration depth $\lambda$ is modified by the electron-phonon coupling constant $\lambda_{ep}$ according to $\lambda^{2}=m_e(1+\lambda_{ep})/(\mu_0e^2\rho_s)$, where $m_e$ is the electron mass, $e$ is the electronic charge, and $\rho_s$ is the superfluid density.
If the superfluid density does not depend strongly on the Ca content, one would expect $\lambda$ to peak near the structural QCP instead \cite{Hashimoto2012,Joshi2020}, because of the stronger electron-phonon coupling there \cite{Goh2015,Yu2015,Cheung2018}. We tentatively attribute the dip in $\lambda$ to a strongly enhanced $\rho_s$ at $x_c$. At present, this opposite trend remains a puzzle. In the sister compound \CaIr, a sudden increase in $\lambda(0)^{-2}$, and hence a decrease in $\lambda(0)$, has also been detected by muon spin
relaxation ($\mu$SR) measurements beyond the pressure-induced structural QCP \cite{Biswas2015}, although a peak-like anomaly is not seen in \CaIr\ under pressure. Such a sensitive change in $\lambda(0)$ is seldom reported in typical BCS superconductors. Therefore, the observations in both Rh and Ir series indicate that  structural instabilities may play an essential role in enhancing superconducting properties in this class of compounds. Further investigations may bring new insights to explain the intriguing behaviour of $\lambda$ in superconducting stannides near a structural QCP.

In summary, we have probed the $T$-$x$ phase diagram of the Remeika series \CaSrRhx\ via critical current measurements. The critical current density is significantly larger for $x=0.9$ samples, the composition at which a putative structural quantum critical point has been identified using other techniques. Some of these techniques require the removal of the superconductivity by a magnetic field, or a long-range extrapolation of the normal state properties to the zero-temperature limit. 
Thus, the singular behaviour of the critical current density at zero field is valuable for revealing the existence of a quantum critical point masked by the superconducting dome. Given that the transport critical current is relatively straightforward to measure, our scheme can have a wider impact in identifying the quantum phase transition in a wide range of tunable quantum materials. 
For a superconductor to be useful for application, it must support a large transport current flow. As such, tremendous efforts have been devoted to the enhancement of the critical current density. Our work, as well as the studies on cuprates and Ce-based heavy fermion systems \cite{Naamneh2014,Talantsev2014,Jung2018}, suggests that a delicate tuning of neighbouring instabilities could be an avenue for this endeavour.
\\
\begin{acknowledgments}
We acknowledge useful discussions with Yajian Hu and Tuson Park. This work is financially supported by the Research Grants Council of Hong Kong (A-CUHK402/19, CUHK 14300117), CUHK Direct Grant (4053461, 4053408, 4053463, 4053410) and Grant-in-Aids for Scientific Research from Japan Society for the Promotion Science (18KK0150 and 20J20353).

\end{acknowledgments}


%

\end{document}